\documentclass[11pt,a4paper]{article}
\pdfoutput=1
\usepackage{jcappub}
\usepackage{aas_macros_JCAP}

\newcommand{\hMpc}{h^{-1}\;{\rm Mpc}}
\newcommand{\hMpcI}{h\;{\rm Mpc}^{-1}}
\newcommand{\hGpc}{h^{-1}{\rm Gpc}}
\newcommand{\hMsun}{h^{-1}M_{\odot}}
\newcommand{\void}{\mathrm{v}}

\title{Testing cosmic geometry without dynamic distortions using voids}

\author[a,b]{Nico Hamaus,}
\author[a,b,c,d,e]{P. M. Sutter,}
\author[a,b]{Guilhem Lavaux}
\author[a,b,f]{\\ and Benjamin D. Wandelt}

\affiliation[a]{Sorbonne Universit\'es, UPMC Univ Paris 06, UMR 7095, Institut d'Astrophysique de Paris, F-75014, Paris, France}
\affiliation[b]{CNRS, UMR 7095, Institut d'Astrophysique de Paris,\\ F-75014, Paris, France}
\affiliation[c]{Center for Cosmology \& AstroParticle Physics, Ohio State University,\\ Columbus, OH 43210, U.S.A.}
\affiliation[d]{INFN - National Institute for Nuclear Physics,\\ via Valerio 2, I-34127 Trieste, Italy}
\affiliation[e]{INAF - Osservatorio Astronomico di Trieste,\\ via Tiepolo 11, 1-34143 Trieste, Italy}
\affiliation[f]{Departments of Physics and Astronomy, University of Illinois at Urbana-Champaign,\\ Urbana, IL 61801, U.S.A.}

\emailAdd{hamaus@iap.fr}
\emailAdd{sutter@iap.fr}
\emailAdd{lavaux@iap.fr}
\emailAdd{wandelt@iap.fr}

\abstract{We propose a novel technique to probe the expansion history of the Universe based on the clustering statistics of cosmic voids. In particular, we compute their two-point statistics in redshift space on the basis of realistic mock galaxy catalogs and apply the Alcock-Paczynski test. In contrast to galaxies, we find void auto-correlations to be marginally affected by peculiar motions, providing a model-independent measure of cosmological parameters without systematics from redshift-space distortions. Because only galaxy-galaxy and void-galaxy correlations have been considered in these types of studies before, the presented method improves both statistical and systematic uncertainties on the product of angular diameter distance and Hubble rate, furnishing the potentially cleanest probe of cosmic geometry available to date.}


\arxivnumber{1409.3580}

\begin{document}
\maketitle

%Comments: \NH{Nico}, \PMS{Paul}, \GL{Guilhem}, \BDW{Ben}

\section{Introduction\label{sec:intro}}
Because General Relativity relates the distribution of matter and energy to the geometry of spacetime via Einstein's field equations, observations of the cosmic expansion history allow us to constrain the material constituents of our Universe. In this manner supernova distance measurements have inferred an exotic form of \emph{dark energy} (respectively a cosmological constant) that appears to dominate the cosmic energy budget today and is responsible for the Universe's accelerated expansion~\cite{Perlmutter1999}. While this result has been confirmed by many other experiments, such as the cosmic microwave background missions WMAP~\cite{Komatsu2001} and PLANCK~\cite{Planck2013}, the true nature of dark energy remains elusive and its equation of state too poorly constrained to rule out subsets of competing models trying to explain its origin.

A particularly elegant method to probe the cosmic expansion history has been proposed by Alcock and Paczynski in 1979~\cite{Alcock1979}. It is based on the cosmological principle, stating that the Universe obeys statistical homogeneity and isotropy. Provided this very reasonable principle holds, any significant anisotropies observed in the cosmos must hence be due to erroneous assumptions on its expanding geometry. Galaxy redshift surveys offer the ideal means to perform a so-called Alcock-Paczynski (AP) test, as they are able to map out the three-dimensional matter distribution of large cosmological volumes. Numerous studies have applied the AP test to the baryon acoustic oscillation (BAO) feature in the two-point statistics of galaxies, yielding tight constraints on the Hubble expansion rate and the angular diameter distance as a function of redshift~(e.g., see~\cite{Okumura2008,Blake2011,Reid2012,Kazin2013,Samushia2014,Sanchez2014,Anderson2014,Song2014,Beutler2014} for some of the most recent measurements). Unfortunately, redshift-space distortions from the peculiar motions of galaxies limit the accuracy of this type of technique, as they generate anisotropies in the clustering statistics of galaxies as well. This effect can be accounted for very well on large scales, where linear theory applies~\cite{Kaiser1987} (e.g., the BAO scale). However, on semi-linear scales and beyond, where the bulk of the data is available, theoretical predictions quickly lack sufficient accuracy and redshift-space distortions must be treated as a systematic contamination to the AP effect~(e.g., see~\cite{Scoccimarro2004,Taruya2010,Reid2011,Seljak2011,Kwan2012,Montanari2012,Yoo2012,Hamaus2012,Marulli2012,Vlah2013,Blazek2014,Okumura2014} for some of the latest theoretical and numerical studies on redshift-space distortions).

An alternative observable to the two-point statistics of galaxies for carrying out an AP test is the shape of stacked voids~\cite{Lavaux2012,Sutter2012b,Sutter2014b}. Voids are the underdense regions in the Universe that occupy the space between sheets, filaments and clusters of galaxies, and hence make up the dominant fraction of its volume. Although taken individually, voids exhibit arbitrary shapes and internal structures, their ensemble average (i.e., their stacked galaxy-density profile) obeys statistical isotropy. In practice, a stacked void is constructed by alignment of the volume-weighted centers of each individual void identified in a galaxy survey, and by histogramming the distribution of galaxies around this center. Thus, it is equivalent to the void-galaxy cross-correlation function and hence a two-point statistic like the galaxy auto-correlation function. However, voids have the advantage of being the least evolved structures in the Universe, which makes their theoretical description more feasible and easier to relate to the initial conditions~(e.g., \cite{Neyrinck2013,Leclercq2013}). Nevertheless, stacked voids are subject to redshift-space distortions, which are inherent to the inferred positions of the galaxies used to construct their profiles~\cite{Dubinski1993,Padilla2005,Ceccarelli2013,Paz2013,Micheletti2014}. In contrast to galaxy auto-correlations, however, which are quadratic in the density of galaxies, void-galaxy cross-correlations merely depend on galaxy density linearly. Nonlinear contributions from redshift-space distortions are therefore less severe in this case, allowing to extend the available range of scales for modeling the observational data. Assuming average spherical symmetry, it is further possible to reconstruct the real-space density profile of stacked voids in a model-independent way~\cite{Pisani2014}.

In this paper we propose a third statistic to be adequate for an AP test: void auto-correlations. In contrast to the other two methods, this statistic is not directly affected by redshift-space distortions at all, since the galaxy density does not enter in it. Peculiar motions of galaxies only affect this estimator indirectly when voids are identified in redshift space. However, the difference in void positions from real space is expected to be small, as many galaxies are needed to define a single void, which diminishes the net displacement of its volume-weighted center. With the help of $N$-body simulations and realistic mock galaxy catalogs, in this study we demonstrate this expectation to hold and advocate the consideration of void auto-correlations as an additional, potentially most pristine way of applying the AP test.

\section{Simulation\label{sec:sim}}
We analyze a large simulation that evolved $2048^3$ cold dark matter particles in a $1\hGpc$ box of a PLANCK cosmology~\cite{Planck2013} with the 2HOT $N$-body code~\cite{Warren2013}. Halo catalogs are created using the ROCKSTAR halo finder~\cite{Behroozi2013} with an overdensity threshold of $\delta=200$ to define virialized objects. In order to generate realistic mock galaxy samples we refer to a standard halo occupation distribution (HOD) model using the code developed by ref.~\cite{Tinker2006}. HOD modeling assigns central and satellite galaxies to each dark matter halo of mass $M$ according to a parametrized distribution. We use the parametrization of refs.~\cite{Zheng2007,Zehavi2011}, where the mean numbers of centrals and satellites per host halo are given by
\begin{eqnarray}
\langle N_\mathrm{cen}(M)\rangle &=& \frac{1}{2}\left[1 + \mathrm{erf}\left(\frac{\log M - \log M_\mathrm{min}}{\sigma_{\log M}}\right)\right]\;,\\
\langle N_\mathrm{sat}(M)\rangle &=& \langle N_\mathrm{cen}(M)\rangle\left(\frac{M-M_0}{M_1'}\right)^\alpha \;.
\end{eqnarray}
Here, $M_\mathrm{min}\simeq2.0\times10^{11}\hMsun$, $\sigma_{\log M}\simeq0.21$, $M_0\simeq6.9\times10^{11}\hMsun$, $M_1'\simeq3.8\times10^{13}\hMsun$, and $\alpha\simeq1.12$ are free parameters adapted to the SDSS DR7~\cite{Zheng2007}. The probability distribution of central galaxies is a nearest-integer distribution (i.e., all halos above a given mass threshold host a central galaxy), and satellites follow Poisson statistics. At redshift $z=0$, this results in a distribution of about $2\times10^7$ galaxies of mean number density $\bar{n}\simeq0.02h^3\;{\rm Mpc}^{-3}$ and a mean separation of roughly $3.7\hMpc$. Central galaxies are assigned peculiar velocities of their host halo, and satellite galaxies are given an additional random velocity drawn from a Maxwell-Boltzmann distribution with a standard deviation that matches the velocity dispersion of their host halo's dark matter particles.

These mock galaxies are further utilized to generate void catalogs using VIDE~\cite{Sutter2014d}, a void identification and examination toolkit based on the ZOBOV~\cite{Neyrinck2008} algorithm, which finds density minima in a Voronoi tessellation of the tracer particles and grows basins around them applying the watershed transform~\cite{Platen2007}. It gives rise to a nested hierarchy of voids and subvoids, and we consider all voids in this hierarchy for our analysis. We prevent basins from merging with each other if the minimum ridge density between them is larger than 0.2 times the mean density. This prevents voids of growing too deep into overdense structures~\cite{Neyrinck2008}. We further define void centers as the mean of each void's particle positions, weighted by their Voronoi cell-volume~\cite{Lavaux2012,Sutter2012a}. The effective void radius $r_\void$ is defined as the radius of a sphere comprising the same volume as the watershed region that delimits the void. We consider the range $6\hMpc \le r_\void \le 150\hMpc$, which amounts to a total number of $\sim7\times10^4$ voids with a mean effective radius of $\bar{r}_\void\simeq14.3\hMpc$ in our catalog at $z=0$.

\section{Method\label{sec:method}}
In cosmological redshift surveys, the observed angles $\Theta$ and redshifts $z$ of galaxies observed on the sky are converted to physical distances using the angular diameter distance $D_A(z)$ and the Hubble rate $H(z)$ via
\begin{eqnarray}
 r_\perp &=& D_A(z)\,\Theta\;, \label{r_per} \\
 r_\parallel &=& cH^{-1}(z)\,z\;, \label{r_par}
\end{eqnarray}
where $r_\perp$ refers to physical distances on the plane of the sky, $r_\parallel$ to the ones along the line of sight and $c$ is the speed of light. The angular diameter distance depends on the integral over the inverse Hubble rate as a function of redshift, as well as on the curvature $\Omega_\mathrm{k}=1-\Omega_\mathrm{m}-\Omega_\Lambda$,
\begin{equation}
 D_A(z) = \frac{c}{H_0\sqrt{-\Omega_\mathrm{k}}}\sin\left(H_0\sqrt{-\Omega_\mathrm{k}}\int_0^z H^{-1}(z')\;\mathrm{d}z'\right)\;,
\end{equation}
while the Hubble rate itself also depends on the matter and energy content in the Universe~\footnote{Here we consider a cosmological constant, but this can straightforwardly be extended to an arbitrary equation of state for dark energy. Moreover, the radiation density $\Omega_\mathrm{r}$ can be neglected at late times.},
\begin{equation}
 H(z) = H_0\sqrt{\Omega_\mathrm{m}(1+z)^3+\Omega_\mathrm{k}(1+z)^2+\Omega_\Lambda}\;.
\end{equation}
It is hence inevitable to assume fiducial values for today's Hubble constant $H_0$, matter density $\Omega_\mathrm{m}$, and dark energy density $\Omega_\Lambda$ in order to construct a three-dimensional map of the distribution of galaxies, and to infer a clustering power spectrum $P(k)$ from it. If these fiducial parameters do not coincide with the true cosmological values, not only will the distances in eqs.~(\ref{r_per}) and~(\ref{r_par}) be incorrect, but also the components of the wave vector $\mathbf{k}$ and the power spectrum $P(k)$. This effect is commonly referred to as the AP test~\cite{Alcock1979}, as it allows to determine the true cosmological parameter values in cases where the geometry of the observed structures is known.

One example for such structures are cosmic voids, as on average they obey isotropy, provided the cosmological principle holds. However, as the galaxies that define a void are subject to peculiar motions, they add a contribution to their observed redshift that is no longer purely caused by cosmological expansion, but also by the Doppler effect. This adds a second term in eq.~(\ref{r_par}), which depends on the line-of-sight galaxy peculiar velocity $v_\parallel$,
\begin{equation}
 r_\parallel+H^{-1}(z)v_\parallel = cH^{-1}(z)\,z\;. \label{rs_par}
\end{equation}
Because $v_\parallel$ is a priori unknown in observations, theoretical models are utilized to relate it to the observed density fluctuations of galaxies. Let us write the density fluctuation of a given tracer as
\begin{equation}
 \delta(\mathbf{r}) = \frac{n(\mathbf{r})}{\bar{n}} - 1\;,
\end{equation}
where $n(\mathbf{r})$ is the spatially varying number density of tracers (e.g., dark matter particles, galaxies, or void centers) with a mean value of $\bar{n}$. After applying the Fourier transform
\begin{equation}
 \delta(\mathbf{k}) = \int\delta(\mathbf{r})e^{-i\mathbf{k}\cdot\mathbf{r}}\mathrm{d}^3r\;,
\end{equation}
the three-dimensional tracer power spectrum $P(\mathbf{k})$ is defined via
\begin{equation}
 (2\pi)^3\delta_\mathrm{D}(\mathbf{k}-\mathbf{k'})P(\mathbf{k}) = \langle\delta(\mathbf{k})\delta^*(\mathbf{k'})\rangle\;, \label{P(k)}
\end{equation}
where $\delta_\mathrm{D}$ denotes the Dirac delta function and the asterisk signifies complex conjugation. For the mock galaxies and void centers in our simulation, we estimate $\delta(\mathbf{r})$ using a cloud-in-cell interpolation of tracer particles on a cubic mesh of $512^3$ grid points. The galaxy positions in redshift space are obtained by interpreting their line-of-sight distance to the observer according to the left-hand side of eq.~(\ref{rs_par}). Applying a Fast Fourier Transform (FFT) algorithm, we then obtain the Fourier modes $\delta(\mathbf{k})$ to estimate auto- and cross-power spectra between any two fields via a discretized version of eq.~(\ref{P(k)}). The inverse FFT of these power spectra provides an estimator for the three-dimensional correlation function $\xi(\mathbf{r})$. Finally, subsequent shell averaging yields the two- and one-dimensional counterparts for these two-point statistics.

A convenient statistic to measure the deviation from isotropy in spatial clustering is the ellipticity of the power spectrum
\begin{equation}
 \epsilon(k) \equiv 2\frac{\int k_\parallel'^2 P(\mathbf{k'})\mathrm{d}^3k'}{\int k_\perp'^2 P(\mathbf{k'})\mathrm{d}^3k'}\;, \label{epsilon}
\end{equation}
where $k_\parallel$ and $k_\perp$ are the Fourier-space analogs to $r_\parallel$ and $r_\perp$ in eqs.~(\ref{r_per}) and~(\ref{r_par}), and hence scale oppositely with $D_\mathrm{A}$ and $H$, respectively. The ellipticity is identical to unity only if $P(\mathbf{k})$ is isotropic, so any deviation of $\epsilon(k)=1$ points at anisotropies. Its definition in eq.~(\ref{epsilon}) is motivated in analogy to the ellipticity of the inertia tensor as given in refs.~\cite{Shandarin2006,Lavaux2010,Sutter2013a} and captures any possible signatures of anisotropy. In order to quantify the sensitivity to geometric distortions, respectively to the AP test, we calculate the Fisher matrix~\cite{Fisher1935,Heavens2009} for the ellipticity,
\begin{equation}
 F_{\alpha\beta}=\frac{1}{2}\mathrm{Tr}\left(\frac{\partial\mathbf{C}}{\partial\theta_\alpha}\mathbf{C}^{-1}\frac{\partial \mathbf{C}}{\partial\theta_\beta}\mathbf{C}^{-1}+\mathbf{C}^{-1}\mathbf{M}_{\alpha\beta}\right) \;. \label{Fisher}
\end{equation}
Here, $C_{ij}=\langle\epsilon(k_i)\epsilon(k_j)\rangle$ are the elements of the covariance matrix $\mathbf{C}$ of $\epsilon(k)$ and the elements of $\mathbf{M}_{\alpha\beta}$ are
\begin{equation}
 M^{ij}_{\alpha\beta} = \frac{\partial\mu_i}{\partial\theta_\alpha}\frac{\partial\mu_j}{\partial\theta_\beta}+\frac{\partial\mu_i}{\partial\theta_\beta}\frac{\partial\mu_j}{\partial\theta_\alpha}\;,
\end{equation}
where $\mu_i=\langle\epsilon(k_i)\rangle$. We only consider the parameters $\theta_1=D_\mathrm{A}$ and $\theta_2=H$, since $\epsilon(k)$ is not sensitive to any parameters that control the normalization and shape of the one-dimensional power spectrum $P(k)$. Because a change in $D_\mathrm{A}$ distorts the observed volume geometrically the same way as a change in $H$ does (see eqs.~(\ref{r_per}) and (\ref{r_par})), the two parameters are degenerate and only their product $D_\mathrm{A}H$ can be constrained from an AP test alone. Finally, the optimal uncertainties $\sigma_\theta$ on the parameters of interest $\theta$ are obtained by inversion of eq.~(\ref{Fisher}), yielding the \emph{Cram\'er-Rao} bound
\begin{equation}
 \sigma_\theta^2 \ge F^{-1}_{\theta\theta}\;. \label{Cramer-Rao}
\end{equation}
For the case of purely geometrical distortions, the combination $D_\mathrm{A}H$ is the only parameter that affects the ellipticity in eq.~(\ref{epsilon}), so the matrix inversions and multiplications in eqs.~(\ref{Fisher} -- \ref{Cramer-Rao}) become trivial. We estimate derivatives of a function $\mathcal{F}(\theta)$ numerically by computing finite differences of $\delta\theta/\theta=1\%$ variations in both $D_\mathrm{A}$ and $H$,
\begin{equation}
 \mathcal{F}'(\theta) \simeq \frac{\mathcal{F}(\theta+\delta\theta) - \mathcal{F}(\theta)}{\delta\theta}\;. \label{diff}
\end{equation}
In practice, changes in $D_\mathrm{A}$ and $H$ are realized by $1\%$-rescalings of all tracer coordinates (in redshift space) in the simulation box along either the $x$- and $y$-axes, or along the $z$-axis, respectively. The void-finding and power spectrum estimation steps are repeated for these rescaled boxes, so that eq.~(\ref{diff}) can be applied to obtain derivatives for the mean $\mu$ of the power spectrum ellipticity $\epsilon$. To achieve sufficiently conservative results, we neglect derivatives of the covariance $\mathbf{C}$ in eq.~(\ref{Fisher}), as it is not clear how well this information can be extracted in realistic observations. Moreover, we have checked that the contribution from the mean of the ellipticity greatly exceeds the one from its covariance in eq.~(\ref{Fisher}).

\section{Results\label{sec:results}}
Anisotropies in the power spectrum can be caused by either geometric or dynamic distortions in the redshift-space coordinates. In this section we separately examine each of these distortions with the help of our numerical simulations.

\subsection{Dynamic distortions}

\begin{figure*}[!t]
\centering
\resizebox{\hsize}{!}{
\includegraphics[trim=0 0 0 0,clip]{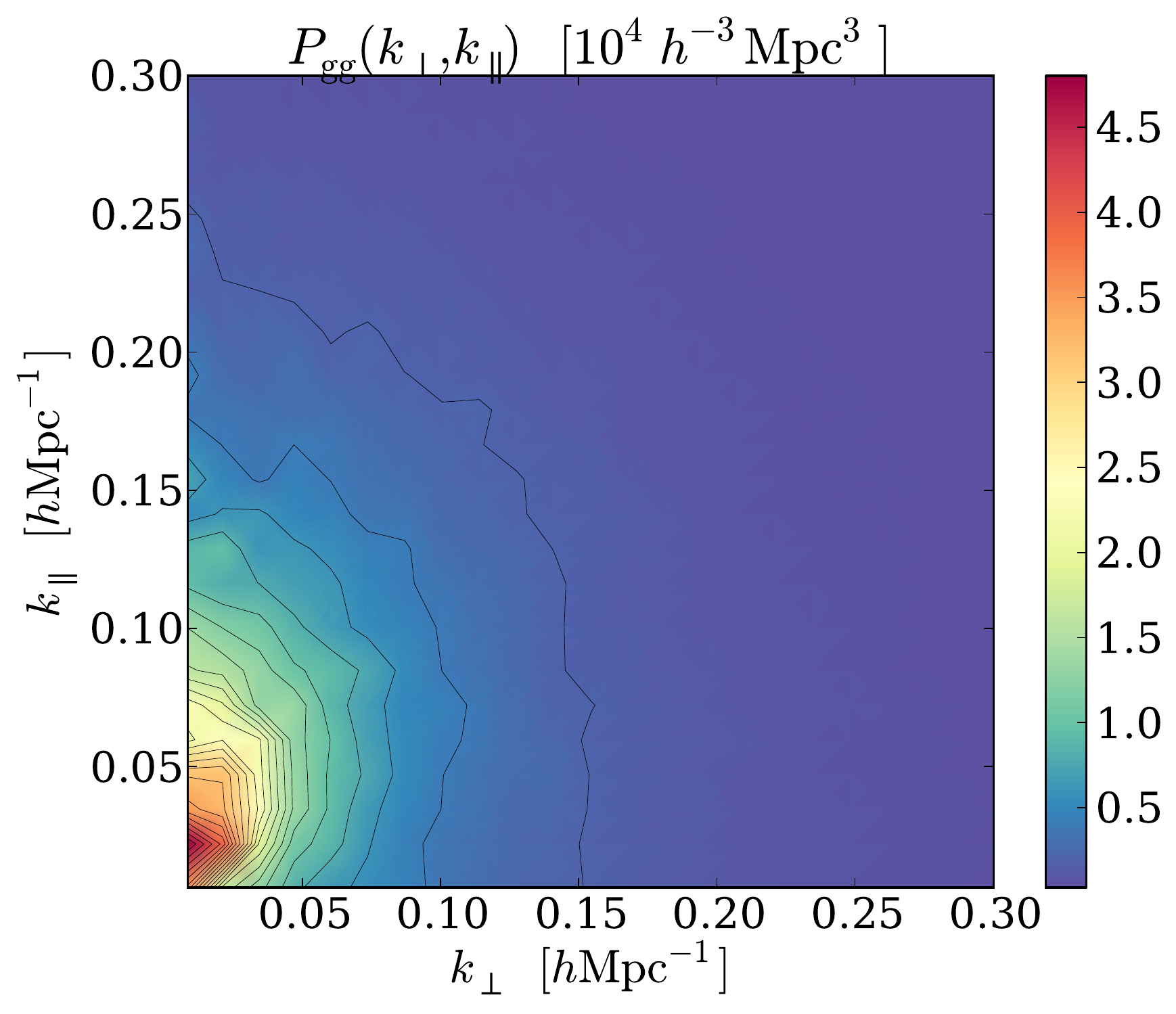}
\includegraphics[trim=0 0 0 0,clip]{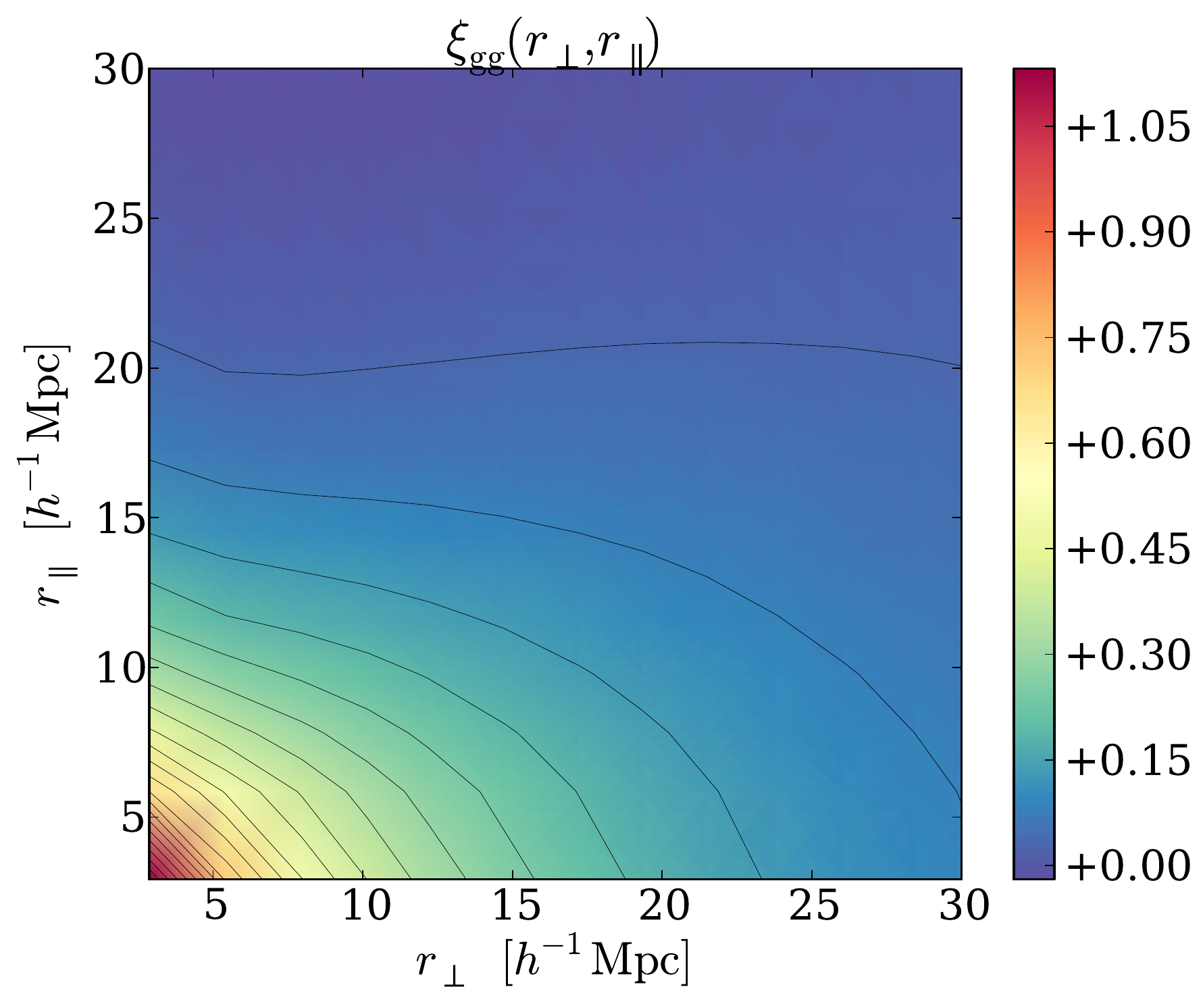}}
\resizebox{\hsize}{!}{
\includegraphics[trim=0 0 0 0,clip]{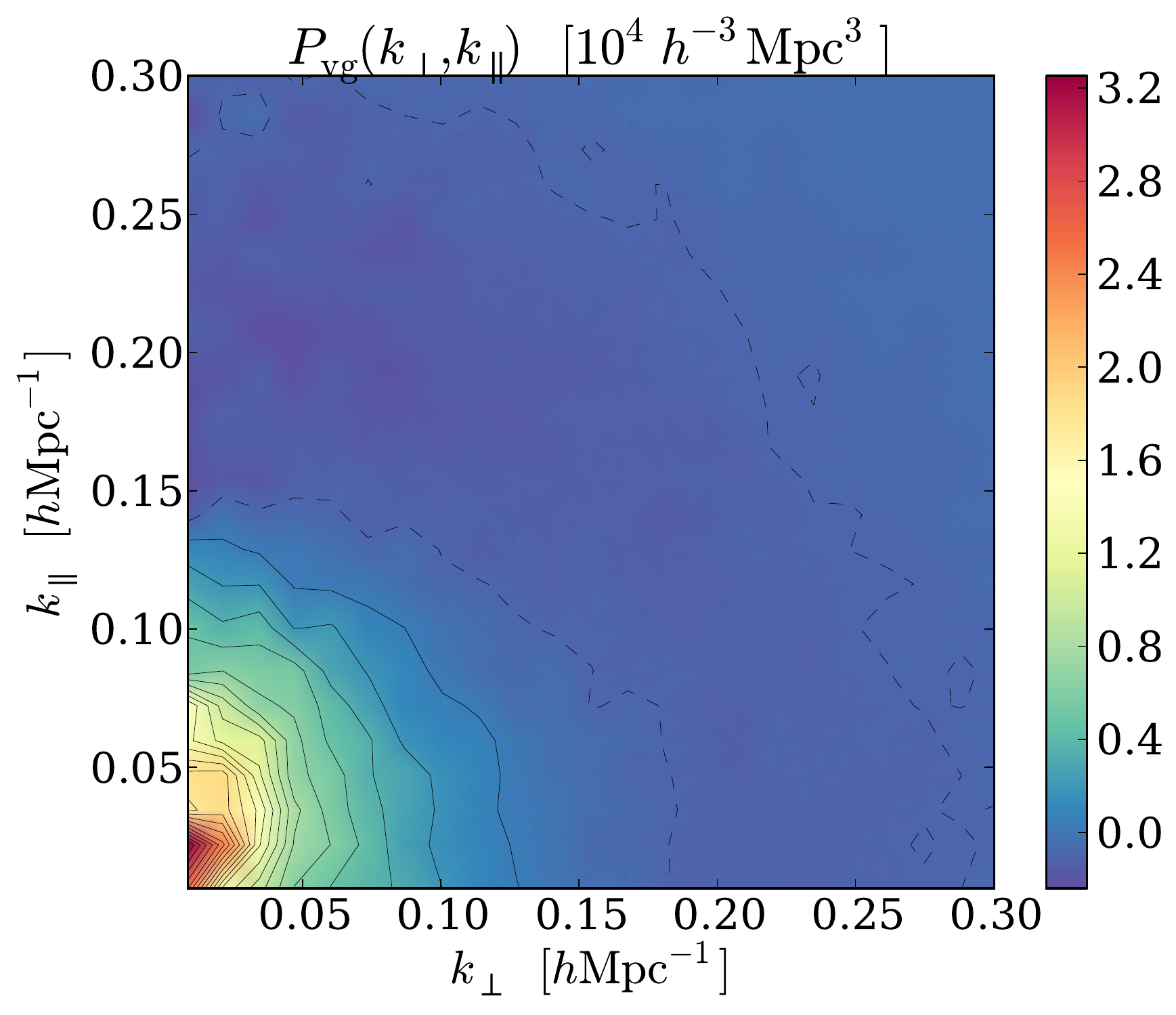}
\includegraphics[trim=0 0 0 0,clip]{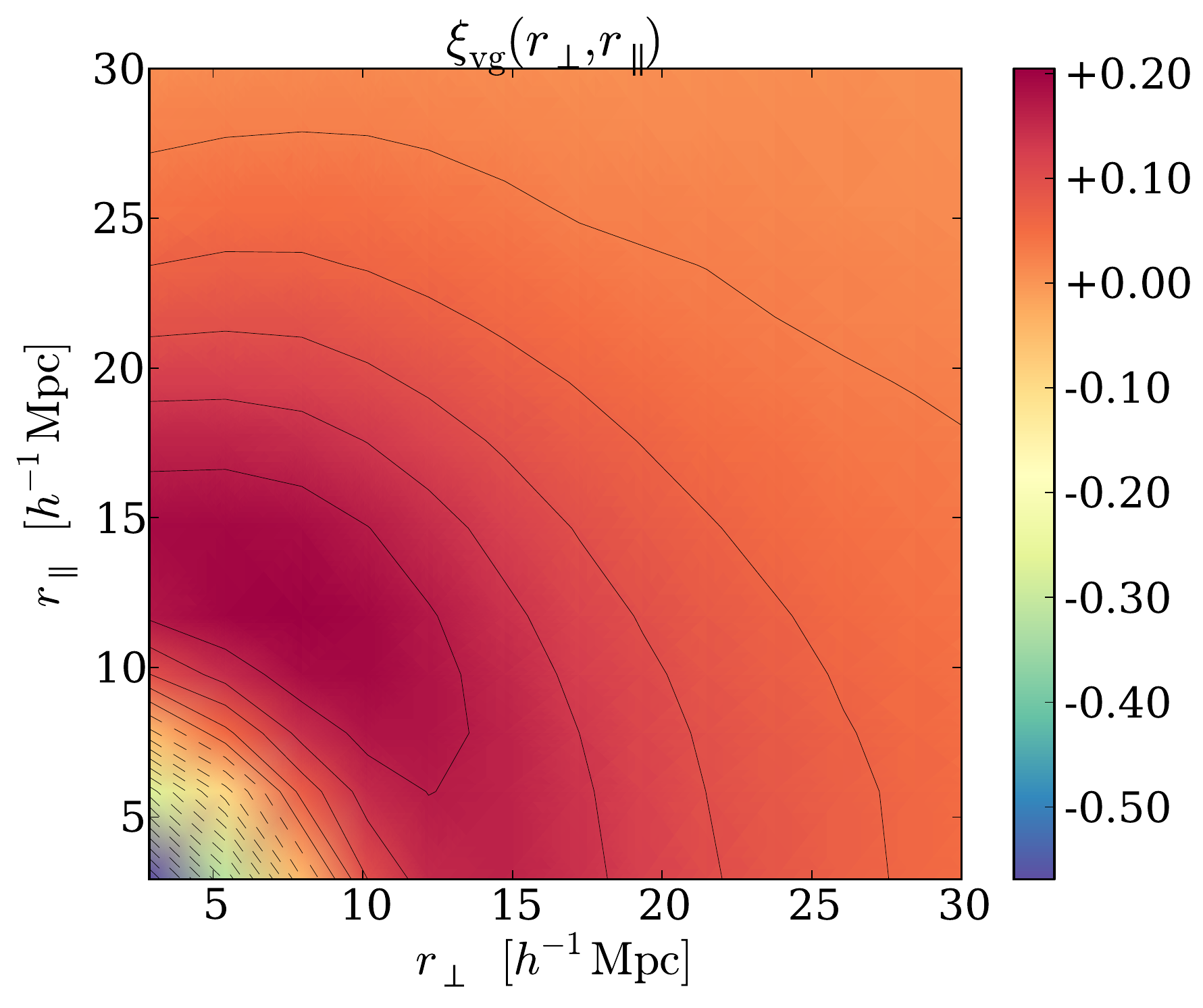}}
\resizebox{\hsize}{!}{
\includegraphics[trim=0 0 0 0,clip]{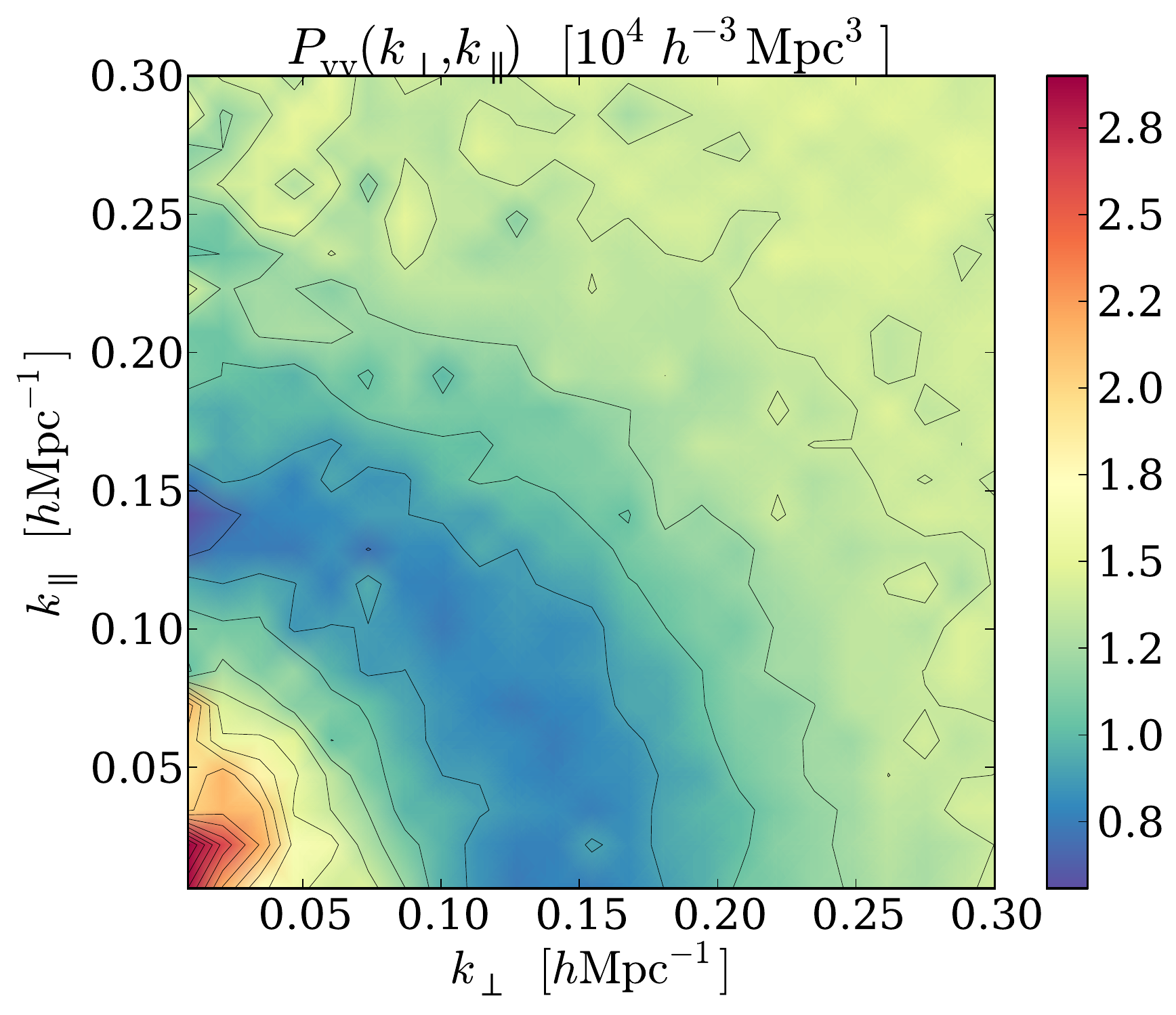}
\includegraphics[trim=0 0 0 0,clip]{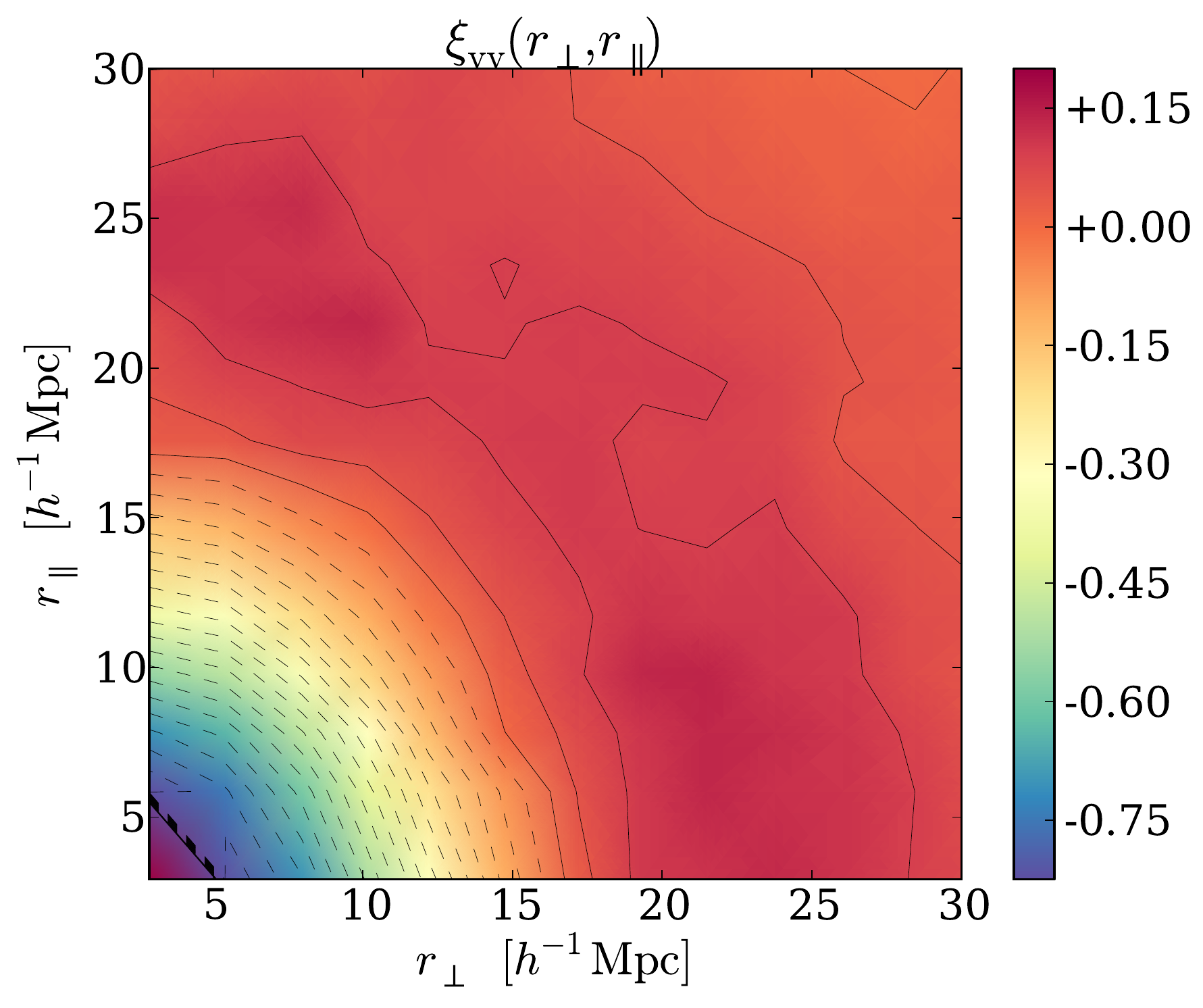}}
\caption{Two-dimensional galaxy auto- (top), void-galaxy cross- (middle) and void auto- (bottom) power spectra (left) and correlation functions (right) in redshift space at $z=0$. Solid lines show positive, dashed lines negative contours.}
\label{fig1}
\end{figure*}

Figure~\ref{fig1} illustrates the impact of purely dynamic distortions on the two-dimensional power spectra and correlation functions of galaxies and voids. Hence, in this case a perfect knowledge of the cosmological parameters (respectively $D_\mathrm{A}H$) is assumed, and anisotropies are only due to redshift-space distortions caused by peculiar motions of galaxies. Their impact is most clearly visible in the auto-correlations of galaxies in both Fourier- and configuration space (top row of figure~\ref{fig1}). On large scales, anisotropies are due to the well-known \emph{Kaiser effect}~\cite{Kaiser1987}: contours are stretched (flattened) along the line of sight in Fourier space (configuration space). This is due to the bulk flow of galaxies towards overdense structures in the linear regime, causing overdensities to be enhanced along the line of sight (``pancakes of god''). On small, nonlinear scales, this behavior is reversed: overdensities become virialized and create random motions of galaxies, which elongate contours of the correlation function along the line of sight (``fingers of god''). This can only marginally be seen in figure~\ref{fig1}, due to its coarse resolution.

In cross-correlations between galaxies and voids (middle row of figure~\ref{fig1}), dynamic distortions are evident as well, but reduced compared to auto-correlations of galaxies. Because the selected sample of voids is overcompensated on average (void-in-cloud effect~\cite{Sheth2004}), on large scales far away from the void center, galaxies feel a net overdensity and hence coherently exhibit infall motions. This causes a Kaiser distortion which is equivalent to the one in galaxy auto-correlations. The most dynamic regions in voids are their compensation walls, which consist of virialized structures such as sheets, filaments and clusters of galaxies. Hence, random motions cause a broadening of contour lines within the compensation walls in configuration space along the line of sight~\cite{Padilla2005,Paz2013,Micheletti2014}. In Fourier space, contours are correspondingly flattened on scales of the compensation wall feature at $k\simeq\pi/\bar{r}_\void$, where the void-galaxy cross-power spectrum changes sign. In the underdense void interior, galaxies exhibit coherent outflow motions, which slightly elongates the contours of the void-galaxy cross-correlation function on small scales along the line of sight~\cite{Paz2013}.

Last but not least, void auto-correlations are shown in the bottom row of figure~\ref{fig1}. Dynamic distortions evidently play a minor role in this statistic, as the contour lines appear to be fairly circular. Residual anisotropies are consistent with being random fluctuations due to the lower number count of voids compared to galaxies, but no systematic trends are manifest. The compensation wall of the void auto-correlation function extends to twice the effective void radius $\bar{r}_\void$, because of mutual void exclusion (see~\cite{Baldauf2013} for similar effects in the halo auto-correlation function). This feature manifests itself as a ring of suppressed power in Fourier space at a scale $k\simeq\pi/2\bar{r}_\void$~\cite{Hamaus2014a,ChuenChan2014}. Existing studies so far have only considered galaxy auto- and void-galaxy cross-correlations for applications of the AP test, but void auto-correlations can be readily obtained from existing data as well~\cite{Padilla2005} and provide additional information on the inferred geometry of large-scale structure. Moreover, systematic uncertainties and biases from redshift-space distortions appear to be negligible in void auto-correlations, which potentially makes this statistic the cleanest one for detecting geometric distortions. We will investigate this conjecture in more detail in the following subsection.

\subsection{Geometric distortions}
\begin{figure*}[!t]
\centering
\resizebox{\hsize}{!}{
\includegraphics[trim=0 0 0 0,clip]{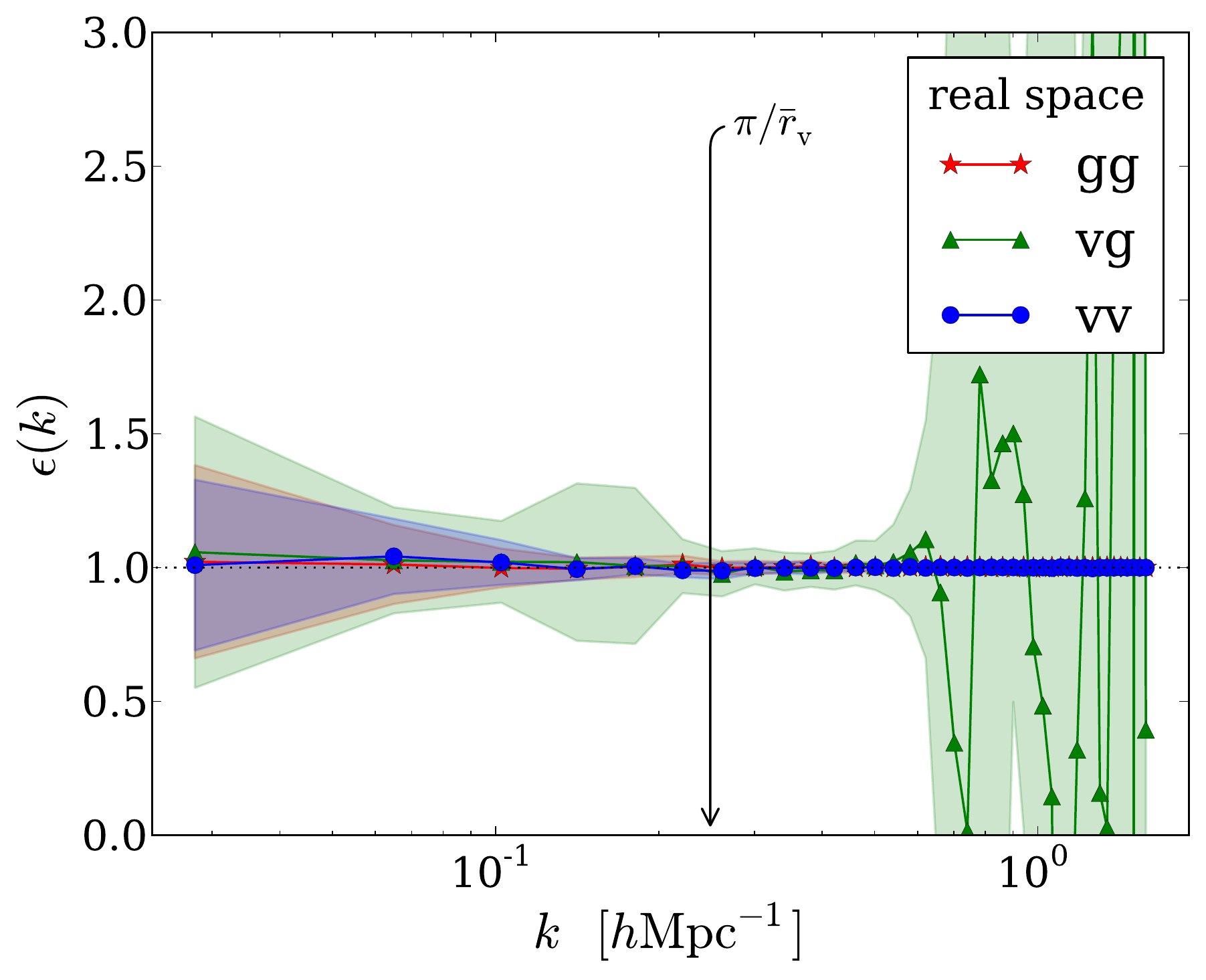}
\includegraphics[trim=0 0 0 0,clip]{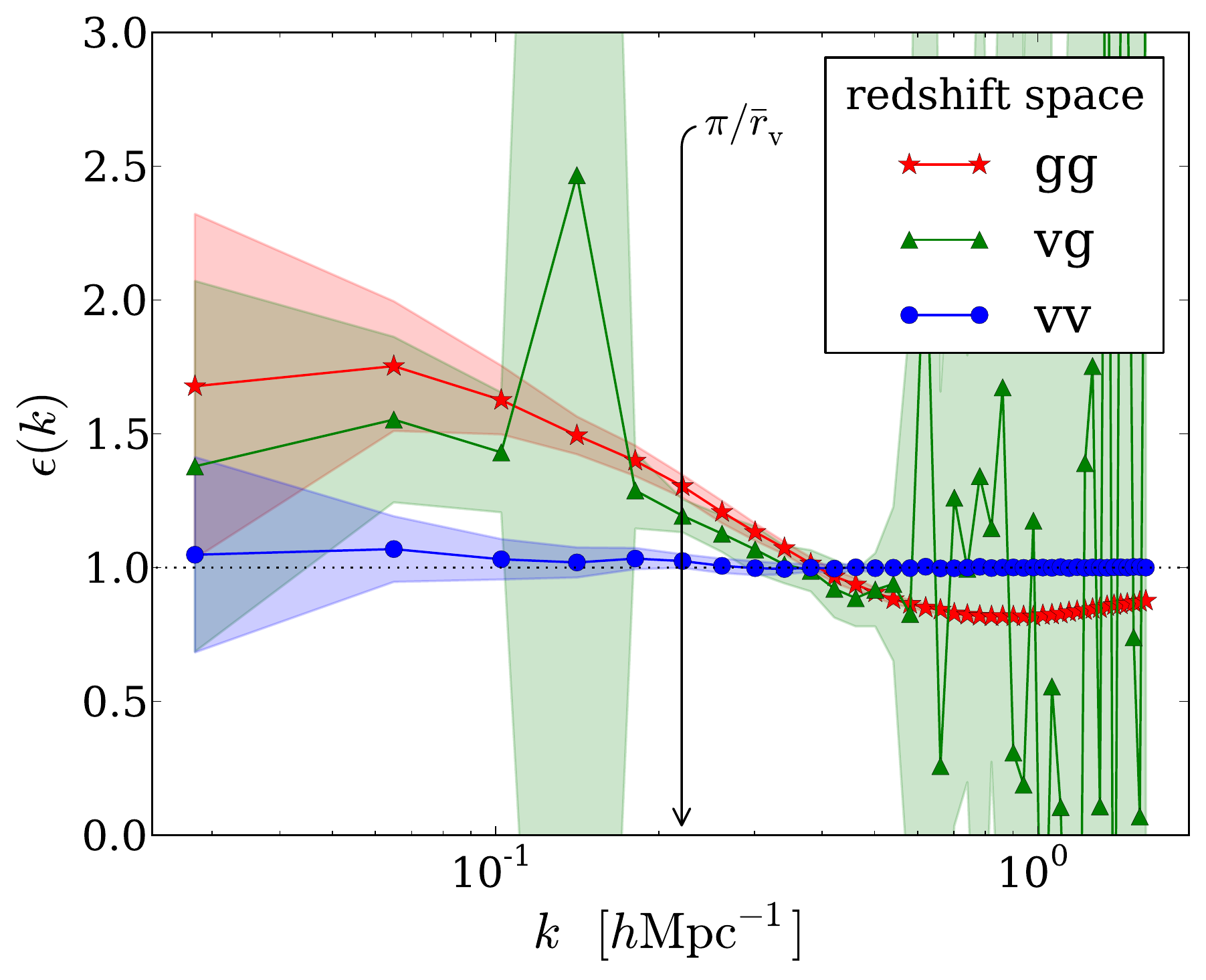}}
\resizebox{\hsize}{!}{
\includegraphics[trim=0 0 0 0,clip]{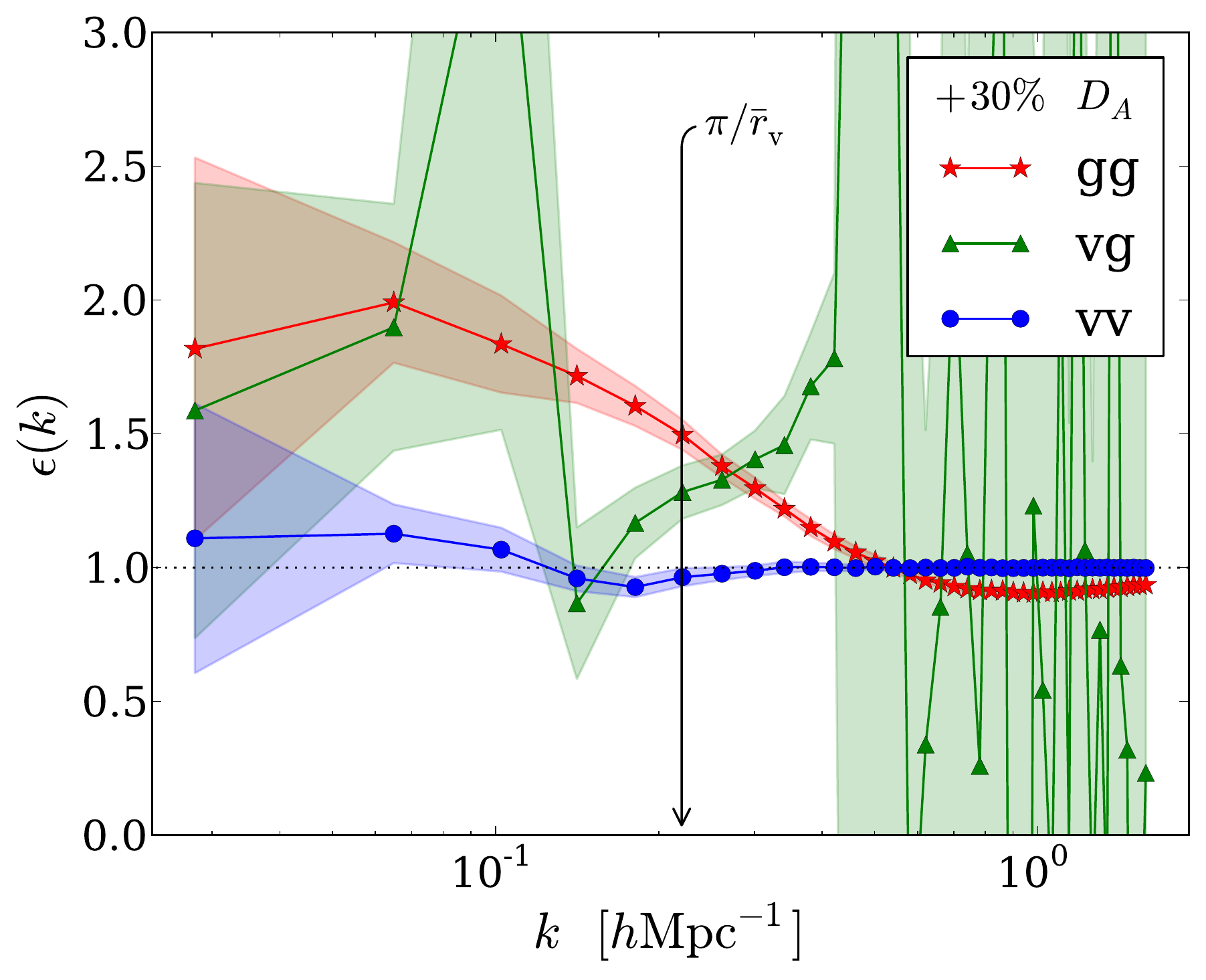}
\includegraphics[trim=0 0 0 0,clip]{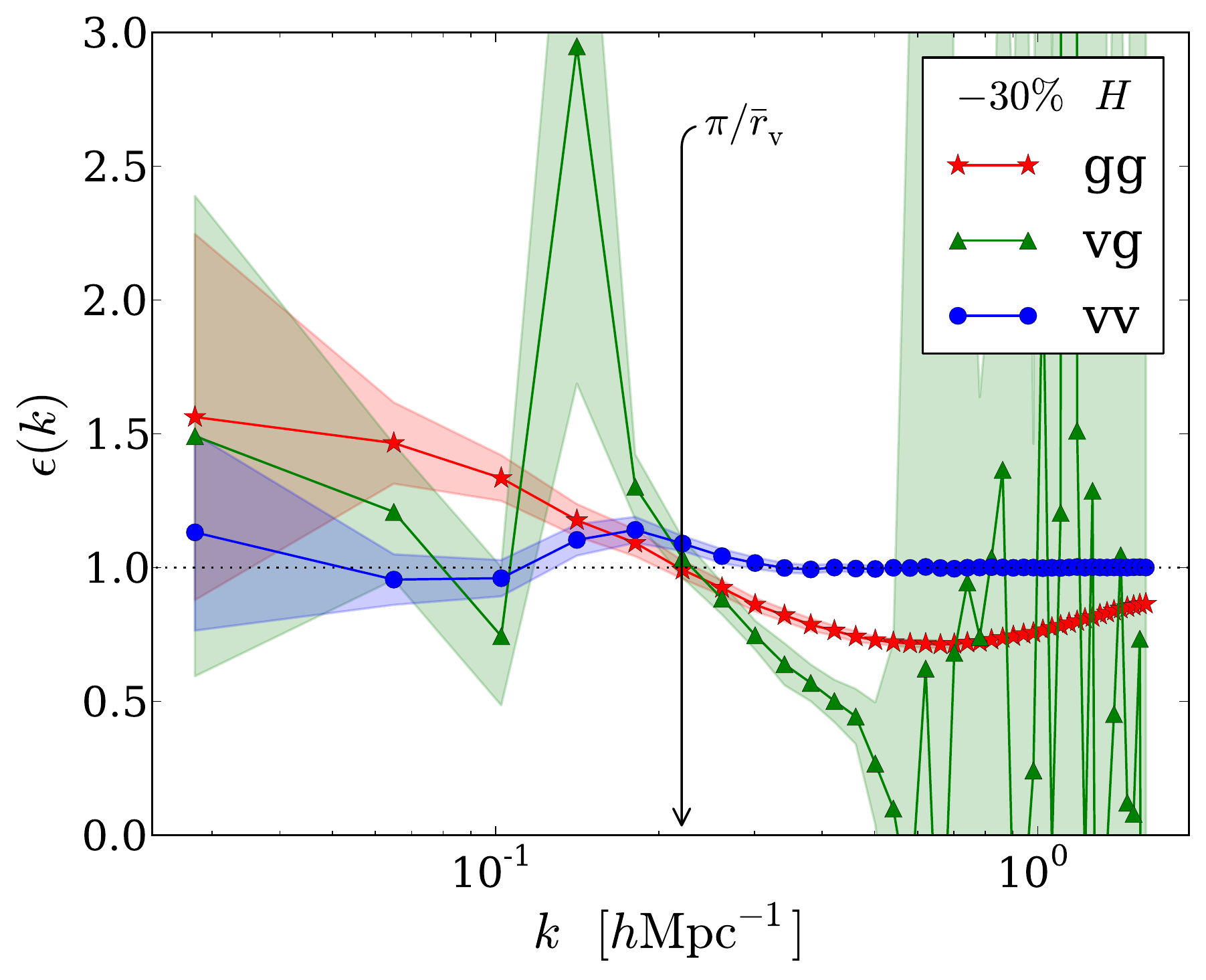}}
\caption{Ellipticity of the three-dimensional galaxy auto- (squares), void-galaxy cross- (stars) and void auto-power spectra (circles) in shells of wavenumber $k$ at $z=0$. Shaded bands show $65\%$ confidence regions obtained from the standard deviation among 27 independent partitions of our simulation box, symbols connected by lines display the corresponding mean values. The top row depicts the fiducial geometry in real (left) and redshift space (right), while the bottom row shows a $30\%$ increase in $D_\mathrm{A}$ (left) and a $30\%$ decrease in $H$ (right) in redshift space. Vertical arrows indicate the average void size of each sample.}
\label{fig2}
\end{figure*}

The ellipticity of the three-dimensional power spectrum quantifies spatial anisotropies in a very concise way, which is why we use it to study the influence of geometric distortions in this section. Figure~\ref{fig2} shows this quantity averaged within shells of increasing wavenumber $k$ for all types of possible correlations between galaxies and voids in different test scenarios. In the top left panel, $\epsilon(k)$ is computed in real space without geometric distortions. This serves as a consistency check, as $\epsilon(k)$ must comply with unity in the case of a statistically isotropic distribution of tracers. Shaded bands show $65\%$ confidence regions obtained from the standard deviation among 27 independent partitions of our simulation box, symbols connected by lines display the corresponding mean values. Evidently, the measured ellipticities do not indicate any deviation from isotropy, their values are all consistent with unity within their uncertainties. Note that relatively high uncertainties in the void-galaxy cross-correlation are due to its partly low power spectrum amplitude, as apparent from figure~\ref{fig1}. In particular, it crosses zero around $k\simeq0.15\hMpcI$ and decays very steeply above $k\gtrsim0.5\hMpcI$ (see also ref.~\cite{Hamaus2014a}).

Switching on peculiar motions of galaxies and hence redshift-space distortions produces the top right panel in figure~\ref{fig2}. For galaxy auto- and void-galaxy cross-correlations the ellipticity of the power spectrum significantly exceeds unity on large scales, a consequence of the Kaiser effect as already observed in figure~\ref{fig1}. On intermediate scales the ellipticity changes sign around $k\simeq0.4\hMpcI$ and remains negative on smaller scales due to the finger-of-god effect caused by virial motions of galaxies. However, for void auto-correlations the ellipticity appears to be remarkably unaffected by redshift-space distortions and remains consistent with unity within the error bars on all scales, as already anticipated from the lower panels of figure~\ref{fig1}.

In the lower two panels of figure~\ref{fig2} we investigate the influence of geometric distortions superimposed on the dynamic distortions from redshift space. For illustration purposes, we chose large perturbations by $30\%$ of opposite sign in $D_\mathrm{A}$ (left) and $H$ (right), as realistic deviations of the percent level would be visible less clearly. Furthermore, instead of rescaling the galaxy positions and repeating the void identification on the rescaled box, we simply shifted the void center coordinates in this case, since these large coordinate perturbations generally destroy the topology of the simulated large-scale structure, which causes voids to fragment. Only in the limit of small perturbations, which we will consider in the quantitative error analysis of the next section, the topology is preserved and ensures a robust void identification.

As apparent from figure~\ref{fig2}, geometric distortions influence the ellipticity of all possible correlations between galaxies and voids. In order to understand how the changes relate to coordinate transformations, let us consider the differential of eq.~(\ref{epsilon}),
\begin{equation}
 \mathrm{d}\epsilon(k) =
 %\frac{\int k_\parallel'^2\mathrm{d}P(\mathbf{k'})\mathrm{d}^3k' \int k_\perp'^2P(\mathbf{k'})\mathrm{d}^3k' - \int k_\perp'^2\mathrm{d}P(\mathbf{k'})\mathrm{d}^3k' \int k_\parallel'^2P(\mathbf{k'})\mathrm{d}^3k'}{\left[\int k_\perp'^2\mathrm{d}P(\mathbf{k'})\mathrm{d}^3k'\right]^2} =
 \frac{\int [2k_\parallel'^2-k_\perp'^2\epsilon(k)]\mathrm{d}P(\mathbf{k'})\mathrm{d}^3k'}{\int k_\perp'^2\mathrm{d}P(\mathbf{k'})\mathrm{d}^3k'}\;, \label{depsilon}
\end{equation}
and let us define the geometric distortion parameters $\alpha_D \equiv D_A/D_A^\mathrm{t}$ and $\alpha_H \equiv H^\mathrm{t}/H$, such that
\begin{equation}
 k_\perp^\mathrm{t} = \alpha_D k_\perp\;,\quad k_\parallel^\mathrm{t} = \alpha_H k_\parallel\;,
\end{equation}
where the superscript ``t'' indicates values in the true underlying, as opposed to the assumed fiducial, cosmology. Using the coordinates $k_\perp$ and $k_\parallel$, we can express the three-dimensional power spectrum in real space and its derivatives with respect to $\alpha_D$ and $\alpha_H$ as
\begin{equation}
 P(\mathbf{k}^\mathrm{t}) = P\left(k^\mathrm{t}(\alpha_D,\alpha_H)\right) = P\left(\sqrt{(\alpha_D k_\perp)^2+(\alpha_H k_\parallel)^2}\right)\;,
\end{equation}
\begin{equation}
 \frac{\mathrm{d}P(\mathbf{k}^\mathrm{t})}{\mathrm{d}\alpha_D} = P'(k)\frac{k_\perp^2}{k}\alpha_D\;,\quad \frac{\mathrm{d}P(\mathbf{k}^\mathrm{t})}{\mathrm{d}\alpha_H} = P'(k)\frac{k_\parallel^2}{k}\alpha_H \;,
\end{equation}
where $P'(k)\equiv\mathrm{d}P(k)/\mathrm{d}k$ and we assume $k^\mathrm{t}\simeq k$. Using these expressions and spherical coordinates with $\mathrm{d}^3k=k^2\mathrm{d}k\,\mathrm{d}\!\cos\theta\,\mathrm{d}\phi$, $k^2=k_\perp^2+k_\parallel^2$, and $\mu \equiv \cos\theta = k_\parallel/k$, the angular and radial integrals in eq.~(\ref{depsilon}) factorize and the derivatives of $\epsilon$ can straight-forwardly be reduced to
\begin{eqnarray}
 \frac{\mathrm{d}\epsilon(k)}{\mathrm{d}\alpha_D} &=& %\alpha_D\frac{\iint_{-1}^1\left[(1-\mu^2)\epsilon(k)-\mu^2\right](1-\mu^2)P'(k')k'^5\mathrm{d}k'\mathrm{d}\mu}{\iint_{-1}^1(1-\mu^2)P(k')k'^4\mathrm{d}k'\mathrm{d}\mu} =
 \frac{4}{5}\alpha_D\left[1/2-\epsilon(k)\right]\frac{\int P'(k')k'^5\mathrm{d}k'}{\int P(k')k'^4\mathrm{d}k'}\;, \label{deps_per} \\
 \frac{\mathrm{d}\epsilon(k)}{\mathrm{d}\alpha_H} &=& %\alpha_H\frac{\iint_{-1}^1\left[(1-\mu^2)\epsilon(k)-\mu^2\right]\mu^2P'(k')k'^5\mathrm{d}k'\mathrm{d}\mu}{\iint_{-1}^1(1-\mu^2)P(k')k'^4\mathrm{d}k'\mathrm{d}\mu} =
 \frac{1}{5}\alpha_H\left[3-\epsilon(k)\right]\frac{\int P'(k')k'^5\mathrm{d}k'}{\int P(k')k'^4\mathrm{d}k'}\;. \label{deps_par}
\end{eqnarray}
Note that the integration over the angle on the sky $\phi$ does not contribute to $\epsilon$, since $P(\mathbf{k})$ does not depend on it. As this calculation is carried out in real space, we have $\epsilon(k)=1$, so the sign and magnitude of changes in $\epsilon$ due to changes in the fiducial $D_A$ or $H$ entirely depend on the amplitude and the slope of the power spectrum. On large scales, $P(k)>0$ and $P'(k)<0$ for all power spectra shown in figure~\ref{fig1}, so $\mathrm{d}\epsilon/\mathrm{d}\alpha_D>0$ and $\mathrm{d}\epsilon/\mathrm{d}\alpha_H<0$, which means an increase in $D_A$ or $H$ increases the ellipticity of the power spectrum (note that $\alpha_D\propto D_A$ and $\alpha_H\propto H^{-1}$). However, when the slope of the power spectrum changes sign, $P'(k)>0$, the ellipticity is decreased. As can be seen from figure~\ref{fig2}, this precisely happens for the void-galaxy cross- and void auto-power spectra beyond their local minima at $k\simeq\pi/\bar{r}_\void$ and $k\simeq\pi/2\bar{r}_\void$, respectively. Note that the amplitude of the void-galaxy cross-power spectrum also changes sign around $k\simeq\pi/\bar{r}_\void$, restoring the large-scale behavior of its ellipticity again.

Although we have neglected the additional effects from redshift-space distortions in the above calculation, the qualitative behavior agrees with our expectations, especially for the void auto-power spectrum, whose redshift-space effects are of minor importance. Therefore, void auto-correlations are potentially the cleanest statistic for carrying out an AP test, as the imprint of geometric distortions can unambiguously be identified. Up to now, existing studies have only focused on galaxy auto- and void-galaxy cross-correlation analyses for that purpose, but void auto-correlations contain additional geometric information that is most likely the least affected by systematic uncertainties. Equations~(\ref{deps_per}) and (\ref{deps_par}) clearly exhibit the fact that the AP test is impossible with a shot noise power spectrum, since $P'(k)=0$ in that case. Thus, only correlated point sets can generate geometric distortions. But even if the slope of the linear power spectrum vanishes, non-linearities can induce $P'(k)\neq0$ on void scales that still allow performing an AP test with void auto-correlations. This is in contrast to the common BAO analysis with galaxies, where the BAO feature is washed out by non-linearities and therefore some information on geometry is lost. In the following section we compare the degree of geometric information that voids and galaxies can provide on a wide range of scales, taking into account the entire shape of their auto- and cross-power spectra.

\subsection{Sensitivity}
Figure~\ref{fig2} suggests the galaxy auto- and the void-galaxy cross-power spectra to be more sensitive to geometric distortions than the void auto-power spectrum, as the shift in amplitude of their ellipticity is larger and affects a wider range of scales. However, not all of this information can be utilized for an AP test, since the anisotropies that arise from redshift-space distortions are poorly understood on nonlinear scales and hence cannot be distinguished from purely geometric distortions. On the other hand, while void auto-correlations do not suffer from this systematic, the relatively low number of voids compared to galaxies reduces the statistical significance of this estimator.

\begin{figure*}[!t]
\centering
\resizebox{\hsize}{!}{
\includegraphics[trim=0 0 0 0,clip]{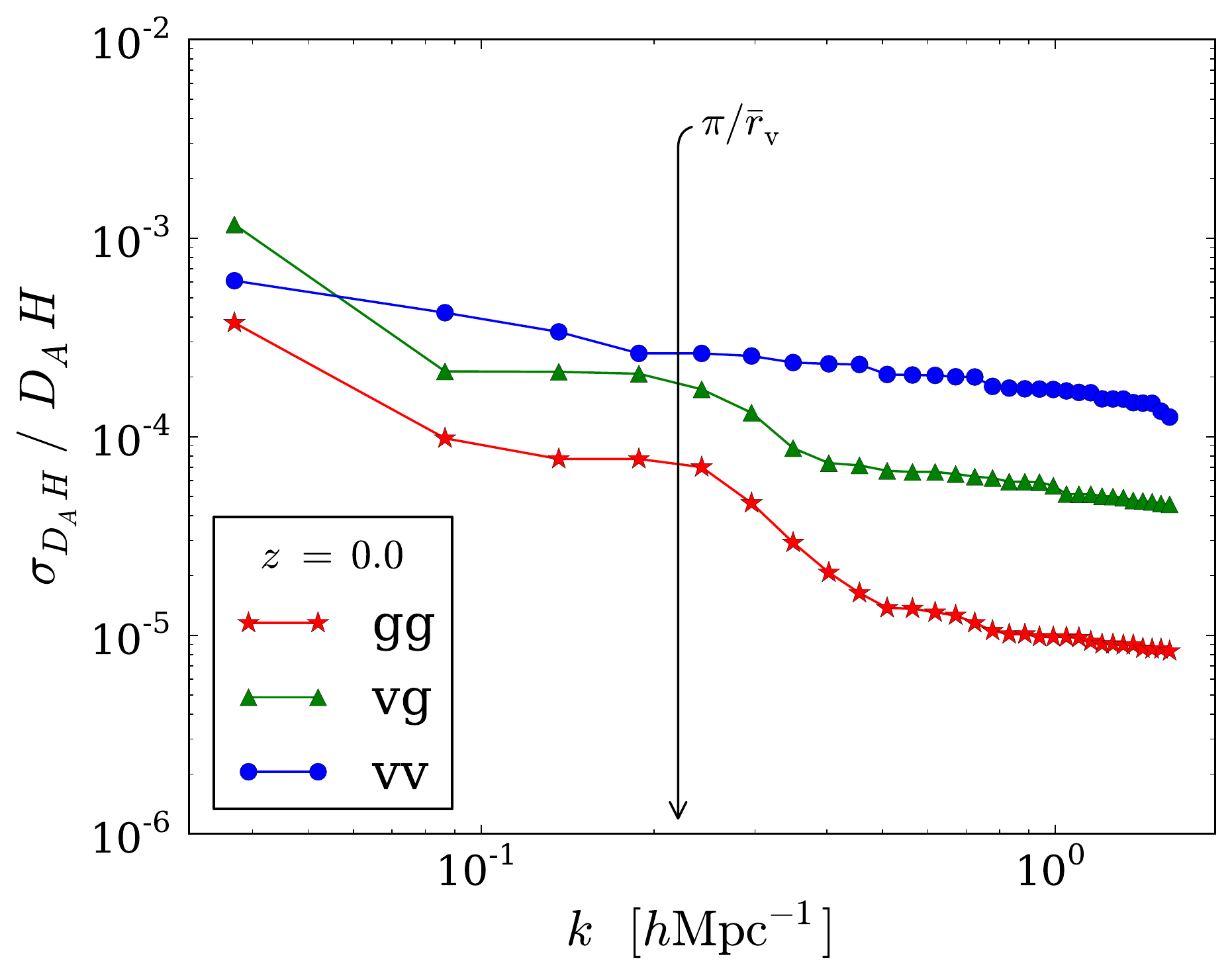}
\includegraphics[trim=0 0 0 0,clip]{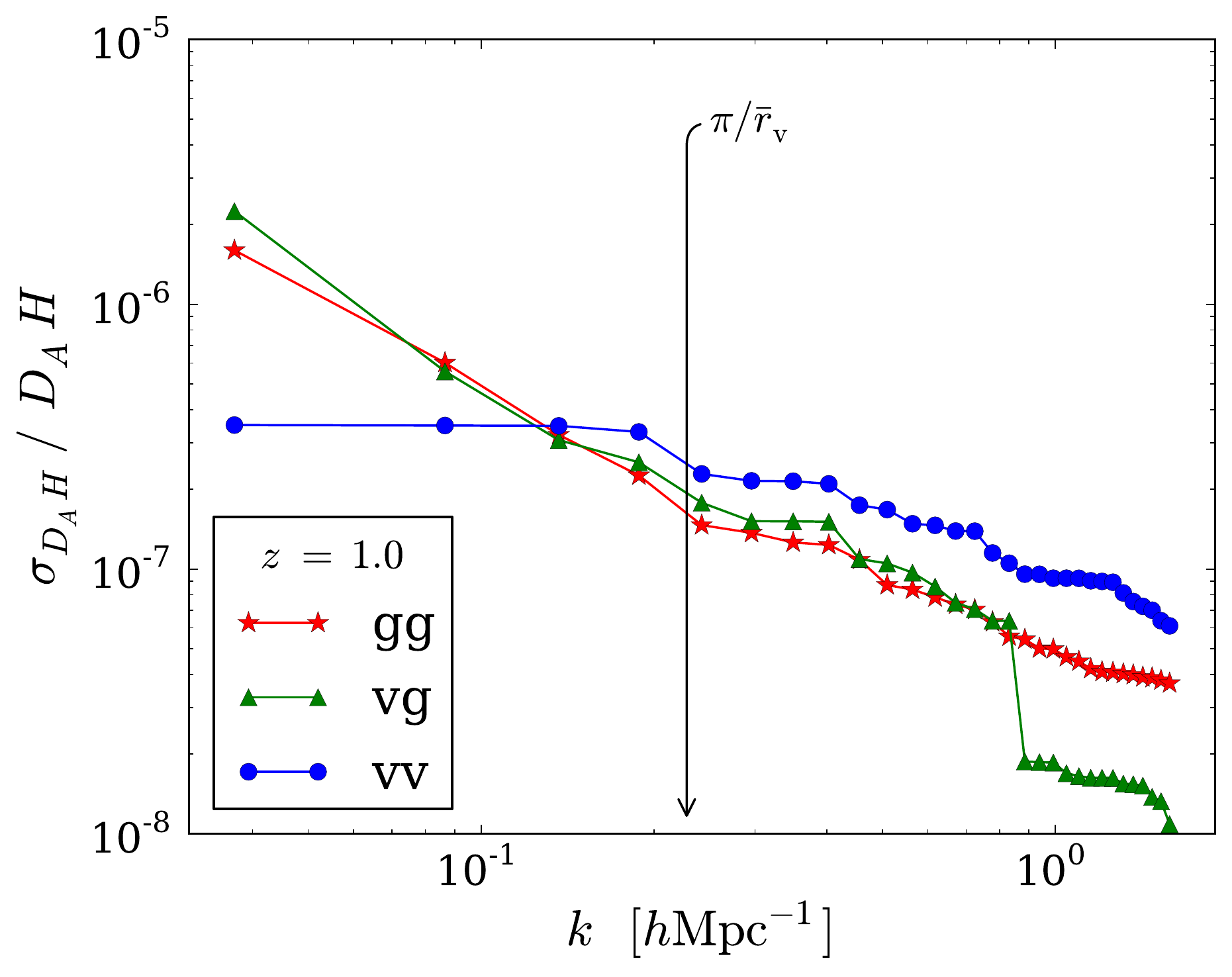}}
\caption{Cumulative $68\%$-uncertainty on $D_\mathrm{A}H$ as a function of maximum wavenumber $k$, as forecasted from a Fisher analysis for the AP test applied to the ellipticity of the three-dimensional galaxy auto- (stars), void-galaxy cross- (triangles) and void auto-power spectrum (circles) from a $1h^{-3}{\rm Gpc}^3$ volume-limited mock galaxy survey. The redshift is stated in the inset of each panel, vertical arrows indicate the average void size.}
\label{fig3}
\end{figure*}

The left-hand panel of figure~\ref{fig3} shows the cumulative $1\sigma$-error on the combination $D_AH$ as a function of increasing Fourier wavenumber $k$, as forecasted via Fisher analysis following section~\ref{sec:method} and based on our $1h^{-3}{\rm Gpc}^3$ simulation volume at redshift $z=0$. The statistical significance on measurements of $D_AH$ from void auto-correlations is less compared to the case where galaxies are taken into account. Its cumulative uncertainty continually decreases up to the characteristic void scale of $k\sim\pi/\bar{r}_\void$ and then gradually saturates due to the domination of shot noise in the void auto-power spectrum~\cite{Hamaus2014a,ChuenChan2014}. Although the uncertainties on $D_AH$ from galaxy auto- and void galaxy cross-power spectra continue decreasing beyond that scale, systematics from redshift-space distortions clearly set a limit on the range of scales that can be used for an AP inference. To the desired level of precision, the common models for galaxy redshift-space power spectra can optimistically reach scales of $k\simeq0.2\hMpcI$ at low redshift. Hence, the final precision on $D_AH$ from either galaxy- or void correlations might individually be within the same order of magnitude.

On the other hand, at higher redshifts the constraints on AP distortions from void auto-correlations become more competitive with the other two cases. As apparent from the right-hand panel of figure~\ref{fig3}, the forecasted precision on $D_AH$ becomes comparable among all three types of correlations at redshift $z=1$, on the largest scales void auto-correlations even yield the best constraints. In a volume-limited sample of galaxies, which we are considering in this case, the abundance of voids increases towards higher redshifts, as many small voids get destroyed via the void-in-cloud effect during cosmic evolution~\cite{Sheth2004,Sutter2014c}. In a fixed observed volume the attainable error on $D_AH$ from two-point correlations roughly scales with the inverse square-root of the total number of tracers~\cite{McDonald2009,Hamaus2012}. This can be understood with the fact that differently biased tracers of the large-scale structure contain complementary information on its underlying density field~\cite{Seljak2009a,Seljak2009b,Hamaus2010,Hamaus2011,Bernstein2011,Abramo2013}. Thus, taking voids into account as additional tracers for redshift-space analyses from galaxy surveys may substantially increase the sensitivity to geometric distortions and hence the accuracy on cosmological parameters.

In order to see whether a specific void selection can increase the AP signal, we have repeated our analysis applying various cuts on different properties of voids, such as effective radius, minimum density, and tree level. However, we find that optimal constraints in terms of Fisher information are achieved when the entire void hierarchy is included. This suggests possible cuts on void catalogs to be as least restrictive as necessary for this type of measurement, but it is possible that very small voids close to the resolution limit are more prone to systematic effects.

\section{Conclusions}
Void auto-correlations are a promising statistic for applications of the AP test and constraining the expansion history of the Universe. In addition to galaxy auto- and void-galaxy cross-correlations, they provide complementary information on cosmological parameters, while at the same time being least affected by systematic effects from redshift-space distortions. Due to the lower abundance of voids compared to the one of galaxies though, the bare statistical constraining power of void auto-correlations is somewhat below that of the other two statistics. Also, complicated survey masks and selection effects in realistic observations have a somewhat more pronounced effect on voids than on galaxies, because voids are extended objects. But the accuracy of void clustering information on small scales compensates for the resulting loss of statistical precision, since it does not suffer from redshift-space systematics.
%In realistic observations that deal with complicated survey masks and selection effects this issue is even more severe, as voids are generally more affected by masking than galaxies, due to their large extent. However, the loss in statistical accuracy from the relatively low abundance of voids can be compensated by their clustering information on small scales, which does not suffer from redshift-space systematics. 
Moreover, other complications that typically interfere with cosmological tests of large-scale structure, such as galaxy biasing, galaxy formation and baryonic effects, do not impose any fundamental limitation on the proposed method, since it only relies on the statistical isotropy of voids.

Redshift-space distortions themselves can be utilized as a powerful tool to constrain cosmological parameters as well, provided their influence on the two-point statistics of galaxies is modeled accurately enough (e.g., \cite{Guzzo2008,Cabre2009,DeLaTorre2013}). However, because of their largely nonlinear nature, this can optimistically be achieved only on scales $k\lesssim0.2\hMpcI$ at low redshifts so far. Void-galaxy cross-correlations are less affected by these nonlinearities, as the galaxy density enters this statistic only linearly and not squared. Therefore, models for the void-galaxy cross-correlation (such as in ref.~\cite{Paz2013}) may be applied on even smaller scales and can utilize many more Fourier modes for cosmological inference. These models require knowledge of the peculiar velocity profile of galaxies around voids, which is not directly observable.

Nevertheless, simulation results suggest density and velocity profiles of voids to be of universal character~\cite{Colberg2005,Sutter2013a,Ricciardelli2014,Hamaus2014b,Nadathur2014}, and their interdependency to be very well described by linear theory down to a few Mpc~\cite{Hamaus2014b}. Simple empirical functions for these profiles can then be used to model the data, their free parameters may either be calibrated to numerical simulations, or be marginalized over. Augmented with void auto-correlations, a full redshift-space analysis of the two-point statistics of galaxies and voids promises to yield the most accurate constraints on the expansion history of the Universe and its cosmological parameters from a given redshift survey. Planned spectroscopic surveys such as Euclid~\cite{EUCLID} or WFIRST~\cite{WFIRST} have the ideal requirements for this type of measurement, but even ongoing experiments like the SDSS~\cite{SDSS}, VIPERS~\cite{VIPERS}, GAMA~\cite{GAMA}, or WiggleZ~\cite{WiggleZ} are expected to benefit from the proposed method. We plan to report on its application to real data in the near future.

\begin{acknowledgments}
We thank Michael Warren for providing his $N$-body simulation and Jonathan Blazek, Adam Hawken, Mark Neyrinck, Nelson Padilla, Alice Pisani, David Weinberg, and Rien van de Weygaert for discussions. This work made in the ILP LABEX (under reference ANR-10-LABX-63) was supported by French state funds managed by the ANR within the Investissements d'Avenir program under reference ANR-11-IDEX-0004-02. This work was also partially supported by NSF AST 09-08693 ARRA. PMS is supported by the INFN IS PD51 ``Indark''. BDW is supported by a senior Excellence Chair by the Agence Nationale de Recherche (ANR-10-CEXC-004-01) and a Chaire Internationale at the Universit\'e Pierre et Marie Curie.
\end{acknowledgments}

\bibliography{ms.bib}
\bibliographystyle{JHEP.bst}

\end{document}